\documentclass[reprint,prl]{revtex4-1}
\usepackage{graphicx}
\usepackage{color}
\usepackage{bm}

\usepackage[caption=false]{subfig}

\def\josaa{J.\ Opt.\ Soc.\ Am. A}
\def\ol{Opt.\ Lett.}
\def\prl{Phys.\ Rev.\ Lett.}
\def\pra{Phys.\ Rev.\ A}

\begin{document}

\title{Complex modes in an open lossless periodic waveguide}
\author{Amgad Abdrabou}
\email{mabdrabou2-c@my.cityu.edu.hk}
\affiliation{Department of Mathematics, City University of Hong Kong, 
  Hong Kong, China}

\author{Ya Yan Lu}
\email{mayylu@cityu.edu.hk}
\affiliation{Department of Mathematics, City University of Hong Kong, 
  Hong Kong, China}

\date{\today}

\begin{abstract}
Guided modes of an open periodic waveguide, with a periodicity in the main propagation
direction, are Bloch modes confined around the waveguide core with no
radiation loss in the transverse directions. Some guided modes can
have  a complex propagation constant, i.e. a complex Bloch wavenumber, even
when the periodic waveguide is lossless (no absorption loss). These
so-called complex modes are physical solutions that can be excited by
incident waves whenever the waveguide has discontinuities and
defects. We show that the complex modes in an open dielectric periodic
waveguide form bands, and the endpoints of the bands can be classified
to a small number of cases, including extrema on dispersion curves of
the regular guided modes, bound states
in the continuum, degenerate complex modes, and special diffraction
solutions with blazing properties. Our study provides an improved 
theoretical understanding on periodic waveguides and a useful guidance
to their practical applications. 
 \end{abstract}

\maketitle


Guided modes of an open optical waveguide, by definition, are confined
around the waveguide core \cite{snyder}. Without absorption and
radiation losses,  
the propagation constant of a guided mode is normally a real number. 
For lossless two-dimensional (2D) waveguides (with 1D refractive index profiles)
that are invariant along 
the waveguide axis, it can be proved that the propagation constant of
any guided mode is real. 
Leaky modes and evanescent modes in the continuous 
spectrum have complex propagation constants, but they 
are not guided modes, since their fields are not confined around the core.
However, it is known that  open lossless 3D waveguides (with 2D
refractive index profiles) can support  
full-vector guided modes with complex propagation  constants 
\cite{jablo94,xie11}. For closed waveguides, these so-called complex
modes are known since the 1960's \cite{mrozo97}. 
 The existence of complex modes is related to the fact that the
 full-vector eigenvalue problem at a fixed frequency with the
 propagation constant being the eigenvalue is non-Hermitian. 
While the complex modes do not carry power along the waveguide
axis, they are physical solutions that can be excited by incident
waves for waveguides with discontinuities or defects, and they cannot
be ignored in any rigorous waveguide analysis when eigenmode
expansions are used. It is also known  that the complex modes are the
cause for numerical instability of the full-vector paraxial beam propagation
method, a classical modeling technique for wave propagation in optical
waveguides \cite{xie11}.   

In this Letter, we show that complex modes also exist in 2D open
lossless periodic waveguides for which the refractive index varies
periodically along the main propagation direction (i.e., the waveguide
axis). To the best of our knowledge, a systematic study of complex modes in 
open periodic waveguides is currently not available. Our results
indicate that complex modes form bands, and for each band, the
propagation constant is a complex-valued function of the real
frequency. We also analyze the endpoints of complex-mode bands. It is
shown that the solution at an endpoint can be 
a regular guided mode with a real propagation constant, a bound state
in the continuum (BIC) \cite{shipman03,port05,mari08,hsu13_2,bulg14b,hu15,yuan17}, a degenerate complex mode, or a diffraction solution
with special blazing properties \cite{popov01}. In the following, we
present theoretical and numerical results 
for complex modes of a 2D periodic waveguide. 

\begin{figure}[h]
  \centering 
\includegraphics[width=0.65\linewidth]{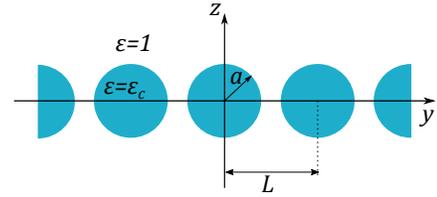} 
\caption{A periodic array of circular cylinders surrounded by
  vacuum.}
\label{1array}
\end{figure}

We consider a lossless 2D structure that is invariant in $x$, periodic
in $y$ with period $L$, and surrounded by vacuum for $|z| > d$. 
A special example  is a periodic array of   
circular cylinders surrounded by vacuum as shown in Fig.~\ref{1array}.
The cylinders are parallel to the $x$ axis and periodically arranged  
along the $y$ axis. The radius and dielectric constant of the
cylinders are $a$ and  $\varepsilon_c$, respectively, and we can let
$d=L/2$. If incident waves  are specified in the surrounding homogeneous media,
the periodic structure serves as a diffractive element, but for waves 
propagating in the $y$ direction, it can be regarded as a 2D periodic
waveguide.  
For simplicity, we consider time-harmonic waves in the $E$
polarization. Thus, the $x$ component of the electric field, denoted as 
$u$, satisfies the following Helmholtz equation
\begin{equation}
  \label{helm}
\partial_y^2 u + \partial_z^2 u + k^2 \varepsilon({\bm  r}) u = 0, 
\end{equation}
where $k$ is the freespace wavenumber, ${\bm  r} = (y,z)$, and 
$\varepsilon({\bm  r})$ is the real dielectric function. 

An eigenmode of the periodic waveguide is given by 
$u({\bm  r}) = e^{ i \beta y} \phi({\bm  r})$, 
where $\beta$ is the propagation constant (or Bloch wavenumber) and 
$\phi$ is periodic in $y$ with period $L$. For a real frequency (i.e.,
a real $k$), if $\phi({\bm  r}) \to 0$ 
exponentially as $z\to \pm \infty$, the eigenmode is a guided mode. A
differential equation for 
$\phi({\bm  r})$ can be easily derived by inserting the Bloch form
into Eq.~(\ref{helm}). This equation for $\phi$, the periodic condition
in $y$, and the boundary condition for $z \to \pm \infty$, give rise
to an eigenvalue problem defined on domain
$\Omega = \{ (y,z):  |y| < L/2, \ |z| < \infty \}$ (one
periodic of the structure), where $\beta$ is the eigenvalue and $k$ is
a real parameter. This eigenvalue problem for guided modes is
non-Hermitian. 

The leaky modes are solutions of a slightly different eigenvalue
problem with an outgoing radiation condition for $z\to \pm \infty$.
If the periodic structure is regarded
as a diffractive element, it is 
useful to regard $\beta$ as a real parameter, $k$ as the eigenvalue,
and impose outgoing radiation condition as $z\to 
\pm \infty$. This leads to a different non-Hermitian eigenvalue
problem with complex-frequency outgoing radiating solutions, i.e., the 
resonant modes (also called resonant states, guided resonances, or quasi-normal
modes) \cite{fan02,amgad19}. It is also useful to consider 
the eigenvalue problem for which $k$ is the (possibly complex) eigenvalue
and $\beta$ is related to $k$ such that $\beta/k$ is real and given
\cite{gras19}.

For a lossless periodic waveguide, guided modes with a real $\beta$
typically exist below the light line, i.e. for  
$k < |\beta|$,  and $\beta$ can be restricted to the interval
$(-\pi/L, \pi/L]$ due to the periodicity. Such a guided mode will be
referred to as a regular guided mode in this Letter. 
For $|z| > d$, the wave field of a regular guided mode can be expanded in
evanescent plane waves as 
\begin{equation}
  \label{plane}
u({\bm  r}) = \sum_{m=-\infty}^\infty c_m^\pm e^{ i \beta_m y + \gamma_m (d
  \mp z)}, \quad \pm z > d, 
\end{equation}
where $\beta_0=\beta$, $\beta_m = \beta + 2\pi m/L$,  and 
$\gamma_m = \sqrt{ \beta_m^2 - k^2}$ for each integer $m$. Some
guided modes with a real $\beta$ can exist above the light line, i.e.,
for $k > |\beta|$, and they are the BICs \cite{mari08,hsu13_2,bulg14b,hu15,yuan17}.  Equation~(\ref{plane}) is still valid for a
BIC, but $c_m^\pm$ must be zero, if the corresponding $\gamma_m$ is
pure imaginary (for $m= 0$ and possibly other integers). 
We are concerned with complex modes, i.e., guided modes with a complex
$\beta$. 
Although there is no absorption loss ($\varepsilon$ is real) 
and no radiation loss ($\phi \to 0$ as $z \to \pm \infty$), the 
propagation constant $\beta$ of a complex mode has a nonzero imaginary
part.  
If the standard complex square root function (with a branch
cut along the negative real axis) is used to define $\gamma_m$,
Eq.~(\ref{plane}) remains valid for complex modes.


Let $u({\bm  r})$ be a complex mode with a propagation constant $\beta
= \beta' + i \beta''$, where $\beta' = \mbox{Re}(\beta)$ and $\beta''
= \mbox{Im}(\beta)$ are the real and
imaginary parts of $\beta$, and $\beta'' \ne 0$. By reciprocity, we
have another complex mode $v({\bm  r})$ with propagation constant
$-\beta$. Since $k$ and $\varepsilon({\bm  r})$ are real,
$\overline{u}$ and $\overline{v}$  (the complex conjugates of $u$ and
$v$),  satisfy the same Helmholtz equation and are also complex
modes. The propagation constants for  
$\overline{u}$ and $\overline{v}$  are 
 $-\overline{\beta}$ and $\overline{\beta}$, respectively. Therefore,
 if $0 < \beta' < \pi/L$, we have four related complex modes 
 $\{ u, \beta \}$, 
 $\{ v, -\beta \}$,
 $\{ \overline{u}, - \overline{\beta} \}$ and $\{
 \overline{v},  \overline{\beta} \}$. 
If $\beta' = 0$,  there are only two distinct propagation constants 
 $\pm i \beta''$. Typically, the corresponding complex modes are
 non-degenerate, then $\overline{u}$ must be proportional to
 $u$. With a proper scaling, we can force $u$ to be a
 real function. Similarly, $v$ can also be scaled as a
real function. The case for $\beta' = \pi/L$ is similar. It is
necessary to regard 
$\beta = \pi/L + i \beta''$ and $-\overline{\beta} = - \pi/L +
i \beta''$ as the same propagation constant. If the corresponding
complex modes are non-degenerate, we can scale $u$ and $v$ as real functions. 
 
Like the regular guided modes, the complex modes form bands. But since $\beta$ is complex
and $k$ is real, it is more convenient to regard $\beta$ as a
complex-valued function of $k$.  Each complex-mode band corresponds
to an interval of $k$ in which $\beta$ is a differentiable function of
$k$.  Multiplying Eq.~(\ref{helm}) by the reciprocal mode
$v$, and integrating on $\Omega$, we can easily derive the following 
formula 
\begin{equation}
  \label{dbdk}
\frac{d \beta }{dk} = 
\frac{ \displaystyle k \int_\Omega \varepsilon({\bm  r}) u v \, d{\bm  r}}
{\displaystyle -i \int_\Omega v \frac{\partial u}{\partial y} \, d{\bm 
    r}}.
\end{equation}
For the special cases with $\beta' = 0$ or $\beta' = \pi/L$,  we know that
$u({\bm  r})$ and $v({\bm  r})$ can 
be scaled as real functions, thus 
$d\beta/dk$ is a pure imaginary number. 
This means that there could be complex-mode bands 
with fixed $\beta' = 0$ or $\beta' = \pi/L$.

The power carried by a guided mode 
is proportional to 
\begin{equation}
  \label{power}
{\cal P}(u) =  \int_{-\infty}^\infty \mbox{Im} \left(
 \overline{u}\frac{\partial 
  u}{\partial y} \right) dz, 
\end{equation}
and it is a constant independent of $y$. For any complex mode, we have 
 ${\cal P}(u) = 0$.
This can be easily proved by 
multiplying $\overline{u}$ to Eq.~(\ref{helm}), integrating on
$\Omega$, and considering the imaginary part. In addition, 
for a complex mode, 
  $ -i  \int_\Omega \overline{u} 
\partial_y u\, d{\bm  r}
=  L {\cal P}(u) = 0$. 
For a regular guided mode, we can assume 
$v = \overline{u}$, but for a complex mode, $v \ne
\overline{u}$, and thus, 
$ \int_\Omega v \partial_y u \, d{\bm  r}$ is not proportional to ${\cal
  P}(u)$ and is typically nonzero. 

To gain a better understanding on the complex modes, we analyze the
conditions for the endpoints of the bands. 
Let $k_*$ be the freespace wavenumber at the end of
a complex-mode band, then  as $k \to k_*$, we have 
$u \to u_*$ and $\beta \to \beta_*$. Assuming $0 \le \mbox{Re}(\beta_*) \le
\pi/L$, the endpoints may be classified as follows. 
\begin{enumerate}
\item $\beta_*$ is real and $k_* < \beta_*$ (below the light line). It is clear that $u_*$ has to be a regular
guided mode. Meanwhile, the 
the propagation constant of $\overline{v}$ (complex conjugate 
of the reciprocal mode) is $\overline{\beta}$ and it also tends to 
$\beta_*$. If the guided mode with freespace wavenumber  $k_*$ and 
propagation constant $\beta_*$ is non-degenerate, then the limit of 
$\overline{v}$ is also $u_*$ (up to a constant), i.e., two complex 
modes $u$ and $\overline{v}$ collapse to one regular guided mode. In 
such case, $\int_\Omega v_* \partial_y  u_* d{\bm  r} = 0$, and $\beta$
(as a function of $k$) has an infinite slope at $k_*$. As we shall see
in the numerical examples below, this type of 
endpoints may appear for both $\beta_* < \pi/L$ and $\beta_* =
\pi/L$. 

\item $\beta_*$ is real and $\beta_* \le k_* < 2\pi/L - \beta_*$ (above
   the light line with one opening radiation channel). In this case, the zeroth
   diffraction channel is open, that is, the $m=0$ terms in Eq.~(\ref{plane})
   are propagating plane waves. If $\beta''$ associated with
   $u({\bm  r})$  
   is positive, then $\gamma_0 \to i \delta_0$ for $\delta_0= (k_*^2 -
   \beta_*^2)^{1/2} > 0 $ as $ k 
   \to k_*$, thus, the $m=0$ terms in Eq.~(\ref{plane}) are incoming
   plane waves. Since it is impossible to sustain a bounded solution
   with incoming waves only, we must have $c_0^+ = c_0^- =
   0$. Therefore, the limit solution $u_*$ must be a BIC. Assuming the BIC is
   non-degenerate, then the two complex modes 
   $u$ and $\overline{v}$ collapse to the same BIC, 
   $\int_\Omega v_* \partial_y 
   u_* d{\bm  r} = 0$, and $d\beta /dk$ is infinite at $k_*$. 
   
\item $\beta_*$ is real and $k_* \ge 2\pi/L - \beta_*$. In this case,
  the  zeroth,  negative first, and probably more radiation channels are
  open. Assuming $\beta'' > 0$ as before, we have $\gamma_0 \to i
  \delta_0$ as $k \to k_*$, but since $\mbox{Im}(\beta_{-1}^2 - k^2)$
  is negative, $\gamma_{-1} \to - i \delta_{-1}$ where $\delta_{-1} = [ k_*^2 -
(\beta_*-2\pi/L)^2]^{1/2} > 0$. Therefore, the limit solution $u_*$
contains an incoming plane wave for $m=0$ and an outgoing plane wave
for $m=-1$, and there is no outgoing wave in the zeroth diffraction
channel and no incoming wave in the negative first diffraction
channel. The existence of diffraction solutions with such a blazing
property is well known \cite{popov01}. 
Since $\overline{\beta}$ also tends to $\beta_*$, the complex mode
$\overline{v}({\bm  r})$ also converges as $k \to k_*$. The limit of
$\overline{v}$ is the reciprocal diffraction solution of $u_*$. Its zeroth
diffraction channel contains only incoming waves and the negative
first diffraction channel contains only outgoing plane waves. Notice
that $u$ and $\overline{v}$ do not collapse to the same solution, and
$d\beta/dk$ can be finite at $k_*$.

\item $\beta_*$ is complex with a nonzero $\mbox{Im}(\beta_*)$. In
  that case, $u_*$ is still a complex mode. Since a band of complex
  modes corresponds to $\beta$ being a differentiable function of $k$,
  we must have $\int_\Omega v_* \partial_y u_* \, d{\bm  r} = 0$, so
  that $d\beta/dk$ is infinite at $k_*$. It appears that this
  condition can only be satisfied when $\mbox{Re}(\beta_*) = 0$ or
  $\pi/L$ with two complex modes $u({\bm  r})$ and $v({\bm  r})$
  converging to the same solution as $k \to k_*$. 
\end{enumerate}

For the three cases 1, 2 and 4, two complex modes, either
$u$ and $\overline{v}$ or
$u$ and 
$v$, coalesce as $k \to k_*$. Therefore, these cases correspond to
exceptional points (EPs) of the non-Hermitian eigenvalue problem for 
guided modes \cite{heiss12,zhen15,kam17,amgad18,amgad19josab}. 

\begin{figure}[h]
   \centering 
\includegraphics[width=0.48\linewidth]{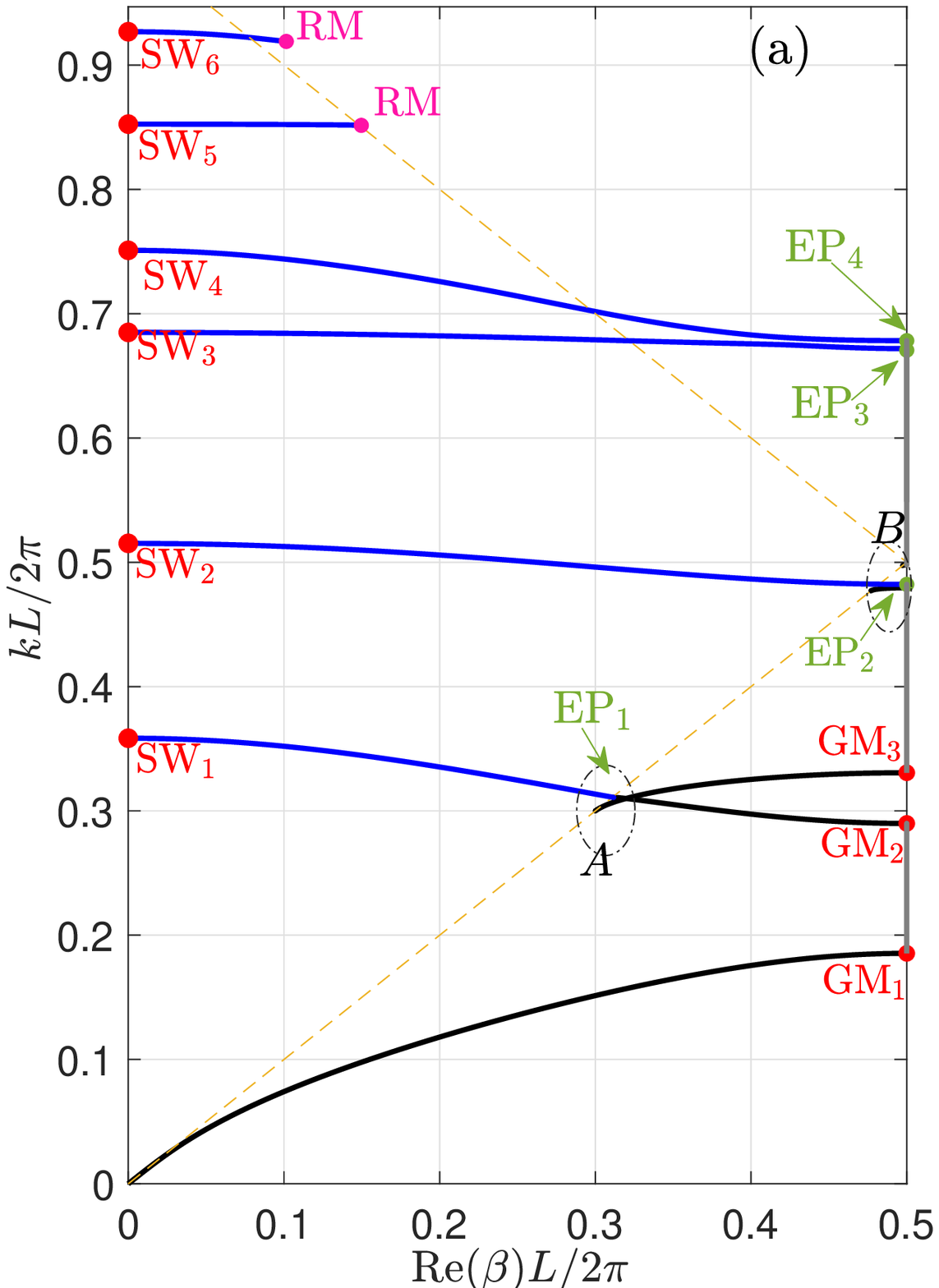} ~\includegraphics[width=0.48\linewidth]{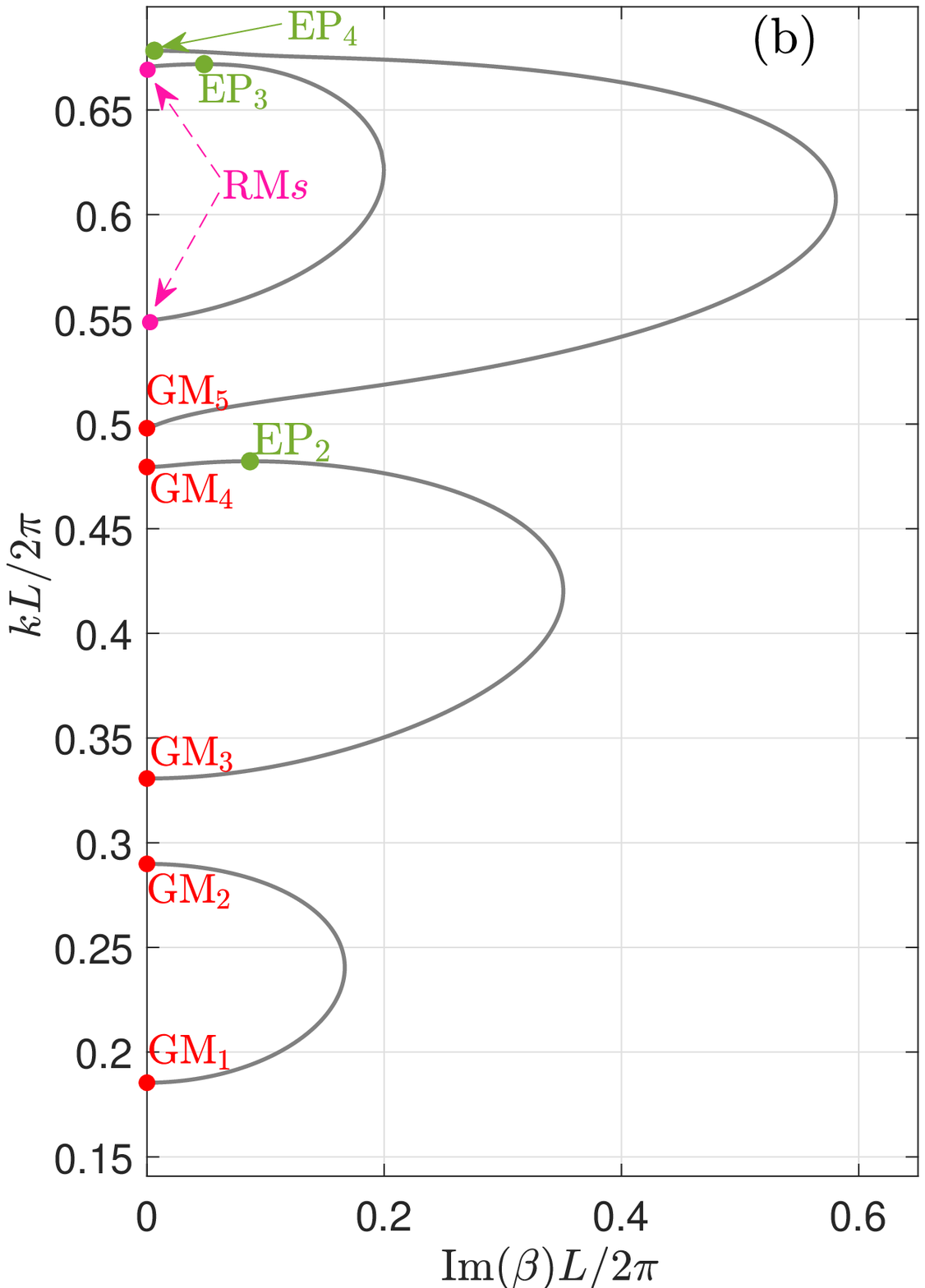}\\  
\includegraphics[width=0.48\linewidth]{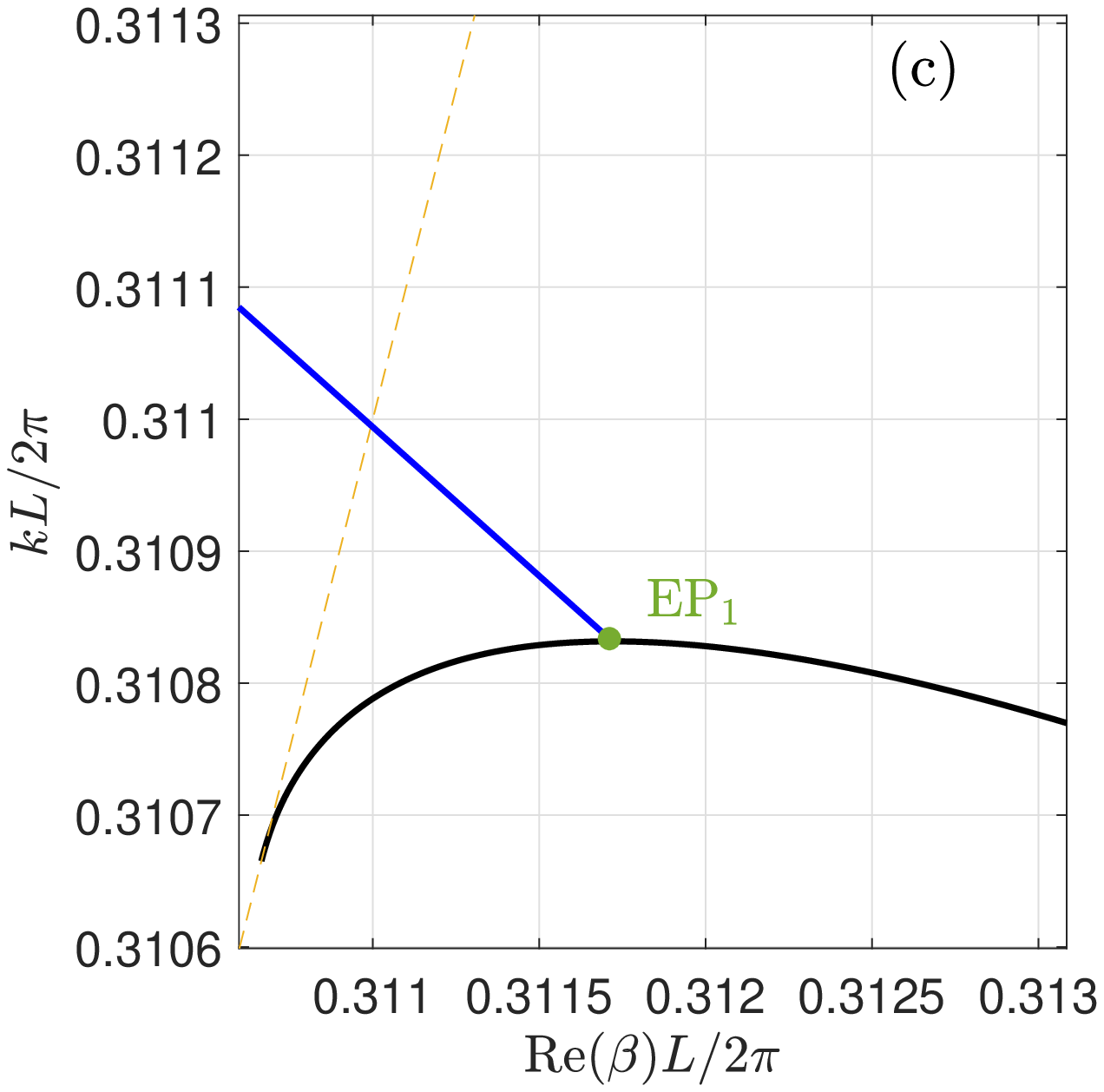} \includegraphics[width=0.48\linewidth]{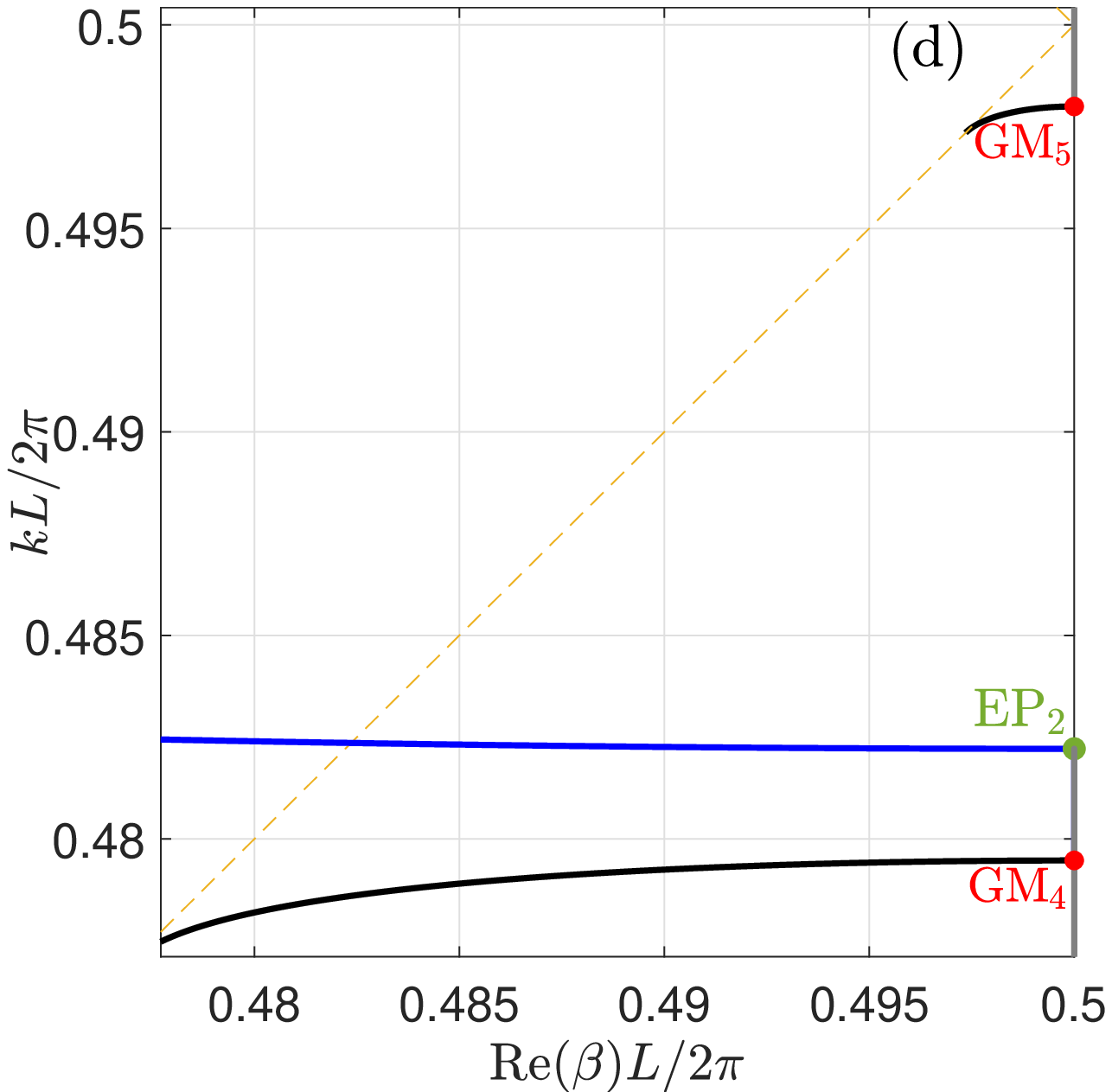}       
\caption{Regular guided modes and complex modes on a periodic array of
  circular cylinders.  (a): Regular guided  modes with a real $\beta$
  (solid black curves) and complex modes for  varying
  $\mbox{Re}(\beta)$ (solid blue curves).  (b) Complex modes for fixed
  $\mbox{Re}(\beta) = \pi/L$. (c) and (d): Zoomed-in plots near points
  $A$ and $B$ in (a), respectively. }
\label{allmodes}
 \end{figure}

For a numerical example, we consider a periodic array of circular
cylinders with radius $a = 0.3L$ and dielectric constant
$\varepsilon_c =15.42$, and show its band structure 
in Fig.~\ref{allmodes}. 
The periodic array has five bands of regular guided modes below the
light line, and they are shown as the solid black curves in
Fig.~\ref{allmodes}(a). The  light line $k = \beta$ is shown 
as the red dashed line 
with a positive slope. 
Since the corresponding intervals for 
$\beta$ are small, the 4th and 5th bands are difficult to see in
Fig.~\ref{allmodes}(a), but they are clearly shown in 
Fig.~\ref{allmodes}(d). 
All five bands start from the light line with a unit slope,
i.e. $dk/d\beta = 1$, and end  with a zero slope at $\beta = \pi/L$. 
The endpoints are marked by ${\sf GM}_j$ for $1 \le j \le 5$ in
Figs.~\ref{allmodes}(a) and (d).  Except for
the second band (with endpoint ${\sf GM}_2$), the dispersion curves of the regular guide modes are increasing
functions of $\beta$ for $\beta \in [0, \pi/L]$. 
In Fig.~\ref{allmodes}(c),  we show the second dispersion curve
near the light line. It is clear that the slope changes signs,  and $k$
(as a function of $\beta$)  has a local maximum at the point marked
as ${\sf EP}_1$.

\begin{figure}[htb]
        \centering 
\includegraphics[width=0.46\linewidth]{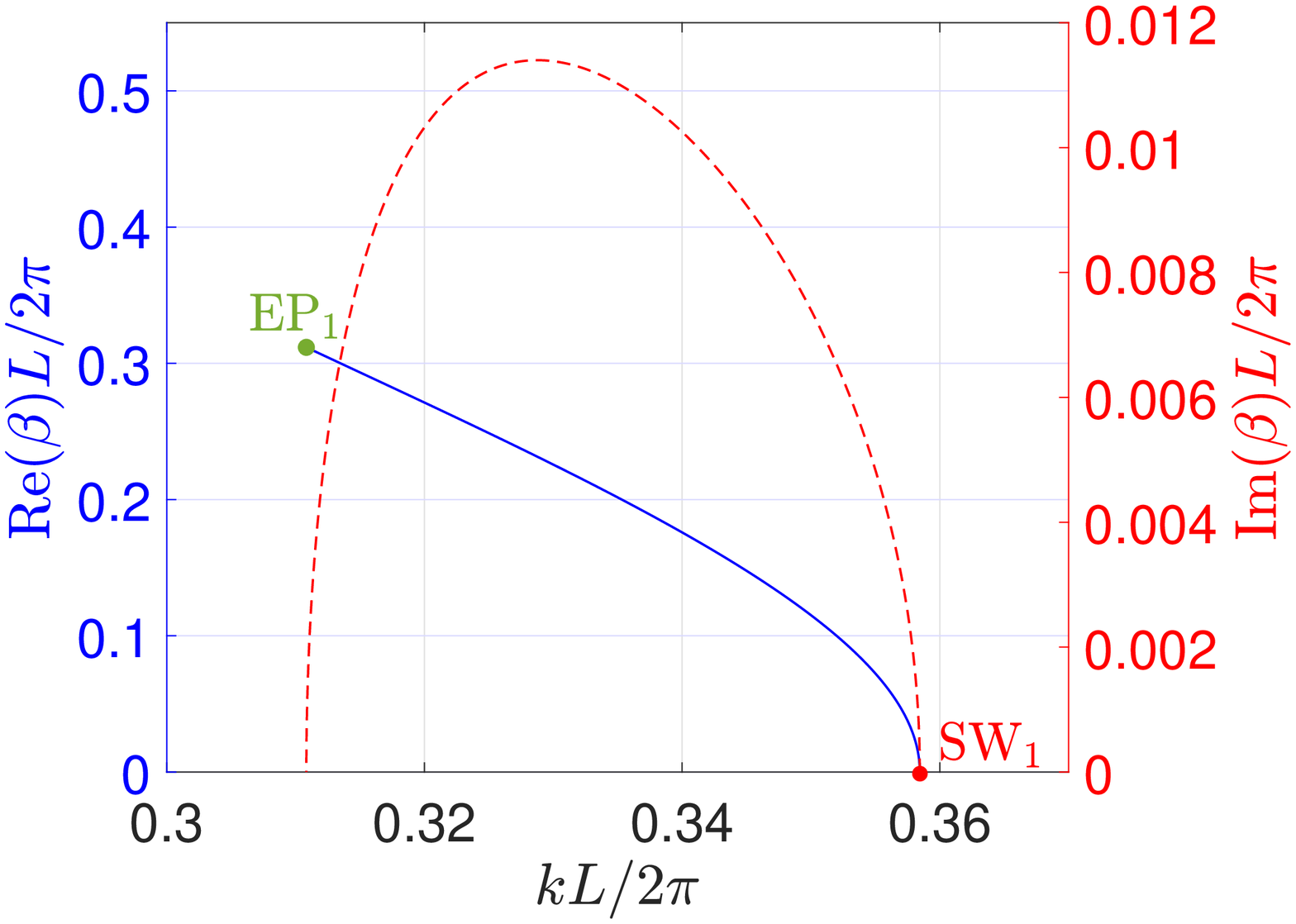}
\includegraphics[width=0.46\linewidth]{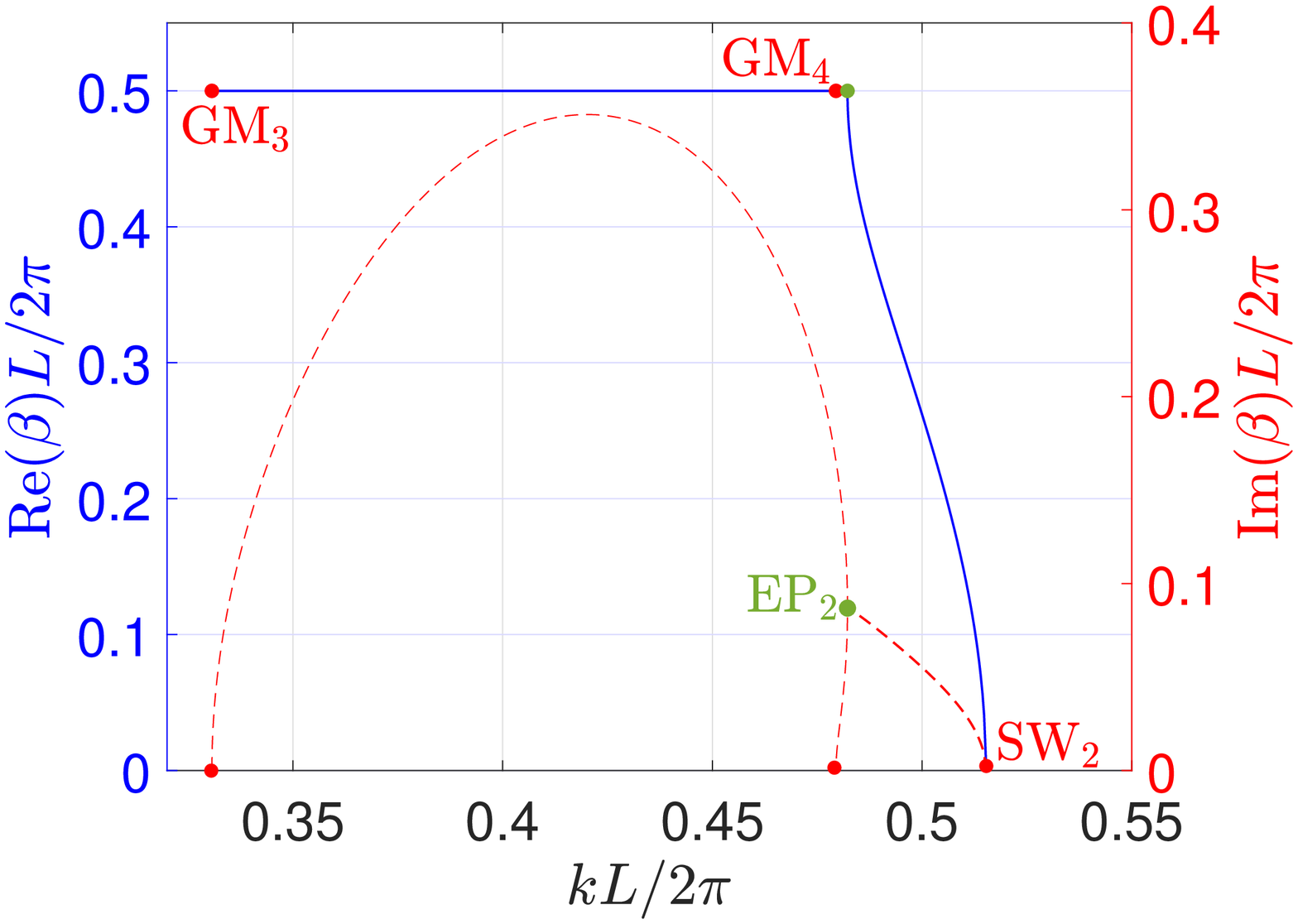}\\
\includegraphics[width=0.46\linewidth]{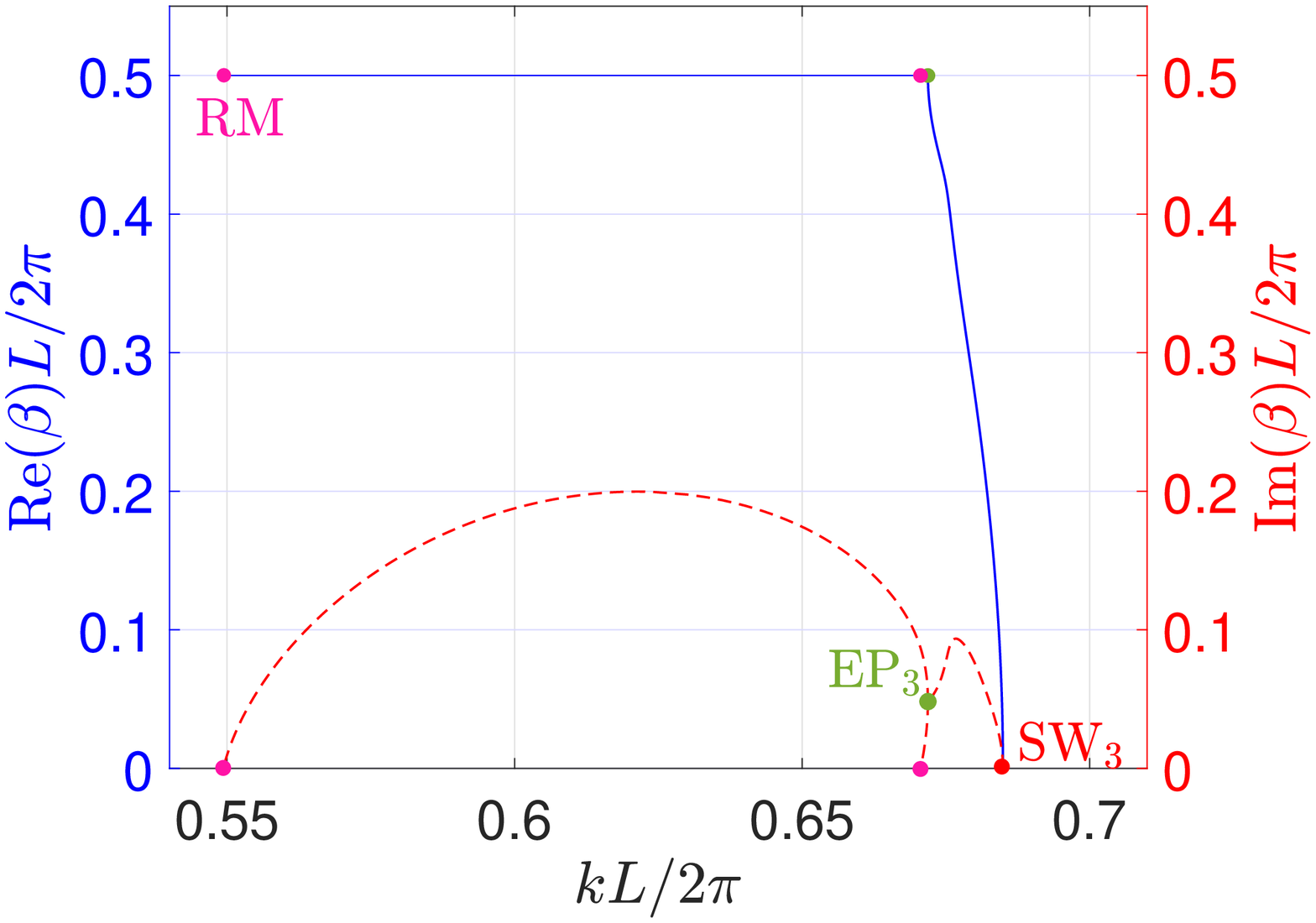}
\includegraphics[width=0.46\linewidth]{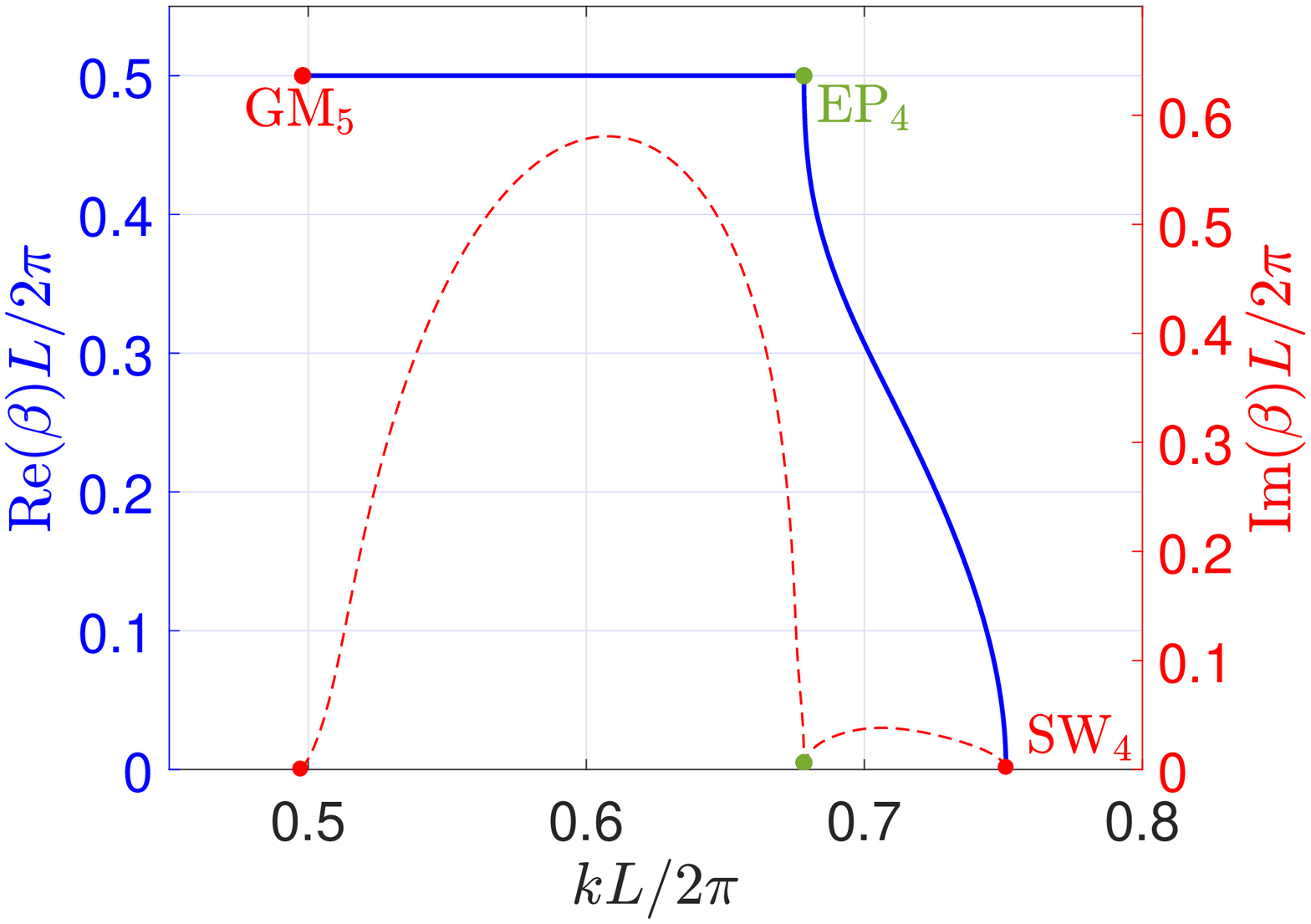}\\
\includegraphics[width=0.46\linewidth]{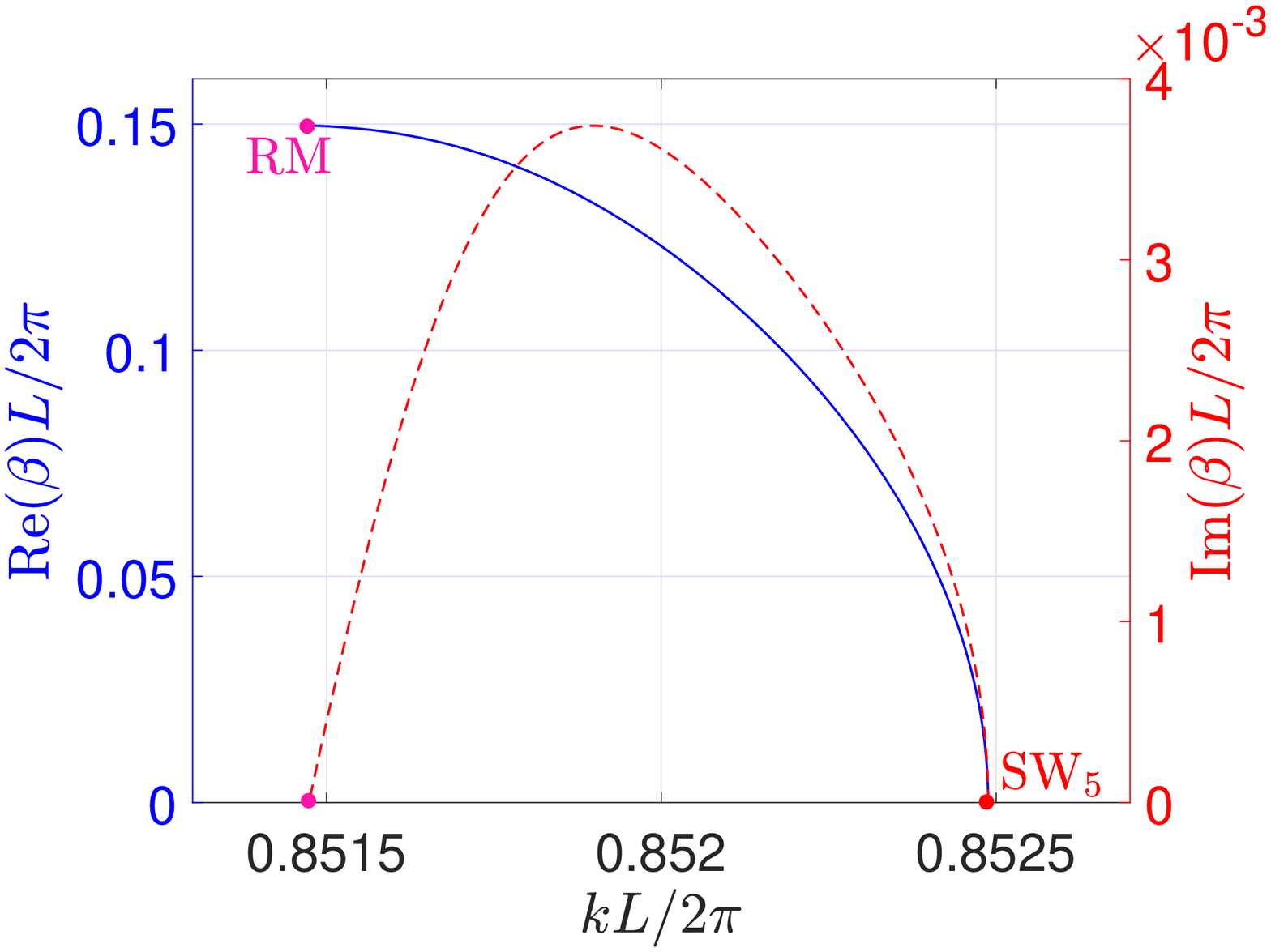}
\includegraphics[width=0.46\linewidth]{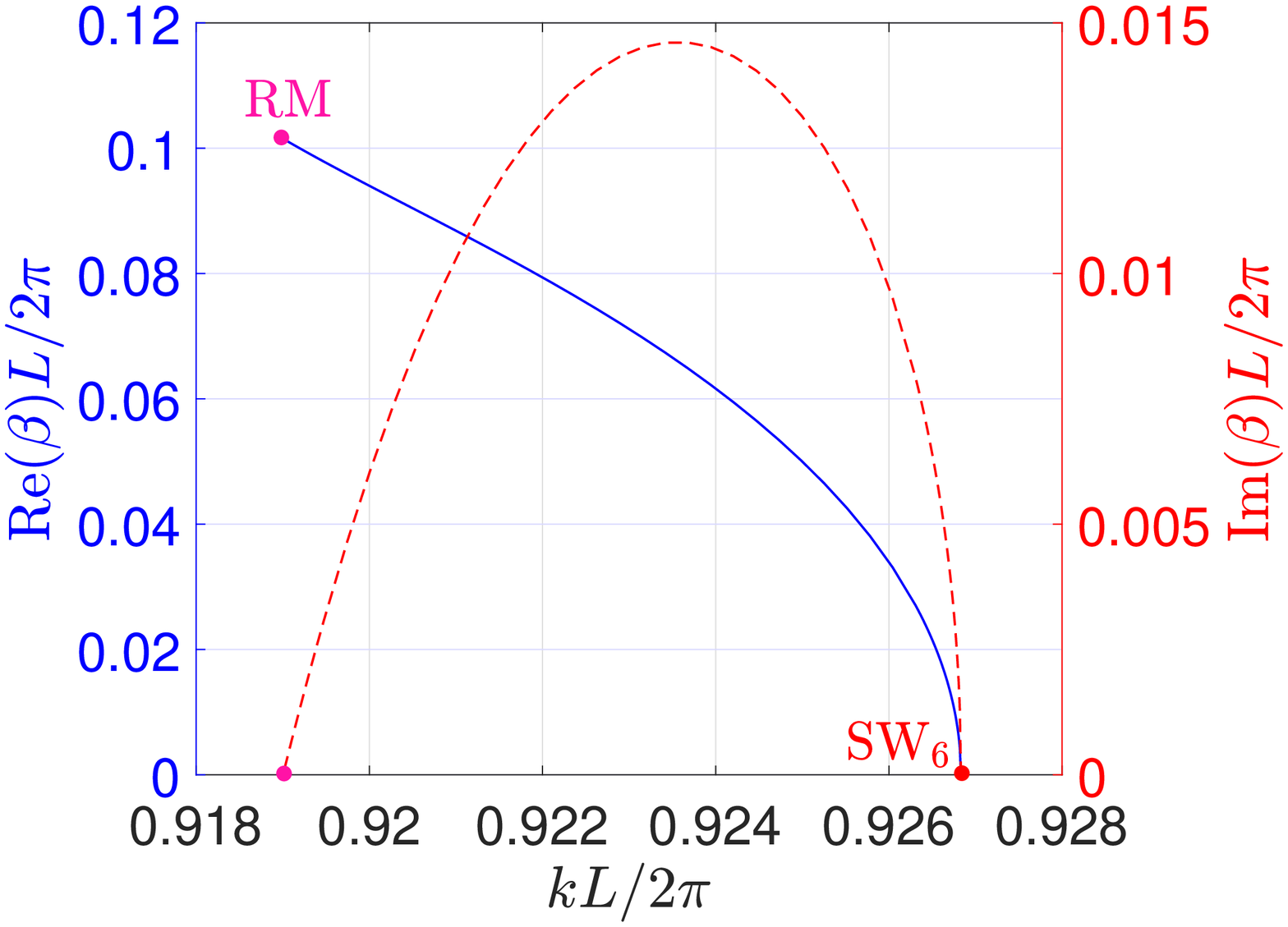}
        \caption{Real and imaginary parts of $\beta$ as functions of $k$ for
          complex-mode bands starting from the standing waves ${\sf
            SW}_j$ for $1\le j \le 6$ in 
          Fig.~\ref{allmodes}(a). The panels with ${\sf
            EP}_j$ ($j=2$, 3, 4) include additional complex-mode
          bands with fixed $\mbox{Re}(\beta) = \pi/L$.}
        \label{allbeta}
\end{figure}

The dispersion curves of the complex modes of this periodic array are
also shown in Fig.~\ref{allmodes}.  
The solid blue curves in Fig.~\ref{allmodes}(a) depict  six
complex-mode bands ($k$ vs. $\mbox{Re}(\beta)$ only) with a
varying $\mbox{Re}(\beta)$. Zoomed-in plots for the first and second
bands near their right
endpoints are shown in Figs.~\ref{allmodes}(c) and (d). Figure~\ref{allmodes}(b) 
shows several complex-mode bands with a fixed $\mbox{Re}(\beta) =
\pi/L$.  In Fig.~\ref{allbeta}, 
we show the real and imaginary parts of $\beta$ as functions of $k$
for the different complex-mode bands. A more detailed description for
these bands are given in the following paragraphs. 

In Fig.~\ref{allmodes}(a), the left endpoints of the six
complex-mode bands are standing waves with a real $k_*$ and $\beta_*
=0$, and they are marked as ${\sf SW}_j$ for $1\le j \le 6$. 
These standing waves are actually symmetry-protected BICs with an 
anti-symmetric (in $y$) electric field. Therefore, all these left endpoints
correspond to case 2, i.e., two complex modes $u$ and 
$\overline{v}$ coalescing to a BIC. 

For all complex-mode bands in Fig.~\ref{allmodes}(a), as $k$ is decreased,
$\mbox{Re}(\beta)$ increases. The right endpoint of the first
complex-mode band is shown as ${\sf EP}_1$  
in Figs.~\ref{allmodes}(a) and (c), and it is 
exactly the local maximum on the second band of
regular guided modes. Clearly, this endpoint corresponds
to case 1, namely, two complex modes $u$ and 
$\overline{v}$ coalescing  to a regular guided mode. 
Let $k_*$ and $\beta_*$ be the freespace wavenumber and propagation
constant of the mode at ${\sf EP}_1$. For $k > k_*$,
to represent a field in the waveguide that decays to zero as $y \to
+\infty$, the two complex modes $u$ and $\overline{u}$ (with
propagation constants $\beta  = \beta' + i \beta''$ and
$-\overline{\beta} = - \beta' + i \beta''$, where $\beta'' > 0$)
should be used in the eigenmode expansion. 
For $k$ slightly less than $k_*$, the periodic array has two guided
modes $u_1$ and $u_2$ 
with propagation constants $\beta_1$ and $\beta_2$ satisfying 
$\beta_1 < \beta_* < \beta_2$. The periodic array also has two
reciprocal modes $v_1$ and 
$v_2$ with propagation constants $-\beta_1$ and $-\beta_2$. 
Since the slopes at $\beta_1$ and
$\beta_2$ have opposite signs, $u_1$ and $u_2$ carry power forward and
backward, respectively. 
To represent a wave field that is outgoing as $y \to +\infty$,  it is
necessary to use $u_1$ and $v_2$ in the eigenmode expansion. 

The right endpoints of
second, third and fourth complex-mode bands shown in
Fig.~\ref{allmodes}(a) have $\mbox{Re}(\beta_*) =
\pi/L$  and $\mbox{Im}(\beta_*) \ne 0$, and they 
are marked as ${\sf EP}_j$ for $2\le j \le 4$ in
Figs.~\ref{allmodes}(a), (b) and (d). In particular, the imaginary part of $\beta_*$
for these three endpoints are shown in Fig.~\ref{allmodes}(b). For
${\sf EP}_4$, $\mbox{Im}(\beta_*)$ is 
small but still positive. Clearly, these endpoints correspond to case
4, i.e., two complex modes $u$ and $v$ merging to a degenerate complex
mode with $\mbox{Re}(\beta_*) = \pi/L$, and $d\beta/dk$ tending
infinity.
For each $j \in \{2, 3, 4\}$, if 
$k_*$ is the freespace wavenumber of the complex mode at ${\sf EP}_j$,
then for $k$ less than $k_*$, two special complex-mode bands
with fixed $\mbox{Re}(\beta) = \pi/L$ emerge. In
Fig.~\ref{allmodes}(b), there are three smooth curves containing 
${\sf EP}_2$, ${\sf EP}_3$ and ${\sf EP}_4$, respectively. On each
curve, the value of $k$ reaches a local maximum (i.e. $k_*$) at ${\sf
  EP}_j$. As $k$ is decreased from $k_*$, two complex modes with 
$\mbox{Im}(\beta) < \mbox{Im}(\beta_*)$ and 
$\mbox{Im}(\beta) > \mbox{Im}(\beta_*)$ (for $k$ close to $k_*$ only)
emerge. 
Since $\beta$ is required to be a differentiable function of $k$ on a
complex-mode band,  each smooth curve containing one
${\sf EP}_j$ in Fig.~\ref{allmodes}(b) corresponds to two
complex-mode bands with $\mbox{Re}(\beta) = \pi/L$. One curve in
Fig.~\ref{allmodes}(b) does not contain ${\sf EP}_j$. Instead, it
connects two regular guided modes ${\sf GM}_1$ and ${\sf
  GM}_2$. Notice that ${\sf GM}_2$ is a local minimum of the second
band of regular guided modes. All other ${\sf GM}_j$ are local maxima
of their corresponding bands.

As shown in Fig.~\ref{allmodes}(b), the six  complex-mode bands
starting from ${\sf EP}_j$ for $2 \le j \le 4$, all end at points on
the line $\mbox{Im}(\beta_*) = 0$. Since these bands have a fixed
$\mbox{Re}(\beta) 
= \pi/L$,  the lower endpoints of these bands all have the
same propagation constant $\beta_* = \pi/L$. It can be observed that
the three bands with a decreasing $\mbox{Im}(\beta)$ (as $k$ is
decreased from that of ${\sf EP}_j$) exist only in 
very small intervals of $k$. For the other three bands,
$\mbox{Im}(\beta)$ initially increases, but eventually decreases
to zero. The lower endpoints of these six bands are either
regular guided modes or special diffraction solutions with blazing
properties. More specifically, the two bands emerging from ${\sf 
  EP}_2$ and one band emerging from ${\sf EP}_4$ end at regular guided
modes ${\sf GM}_3$, ${\sf GM}_4$ and ${\sf GM}_5$, respectively. These
endpoints correspond to case 1 discussed earlier. The two bands
emerging from ${\sf  EP}_3$ end at diffraction solutions marked as 
${\sf RM}_s$ in Fig.~\ref{allmodes}(b). The tiny band to the left
of ${\sf EP}_4$ also ends at a diffraction solution with $\beta_* = \pi/L$. 
All these three diffraction solutions correspond to case 3. Let $u_*$
be any one of these solutions, then $u_*$ has only an incoming
plane wave in the $0$th diffraction order and only an outgoing plane
wave in the $-1$st diffraction order. Since $\beta_0 = \beta_* = \pi/L
= - \beta_{-1}$, the incoming and outgoing waves propagate exactly in
opposite directions.  

Finally,  we consider the 5th and 6th complex-mode bands in 
Fig.~\ref{allmodes}(a). The right endpoints of these two bands are
marked as ${\sf RM}$ and also correspond to case 3. The limiting
solutions at these two endpoints 
are also blazing diffraction solutions that
completely convert the power of the incoming waves in the $0$th
diffraction order to outgoing waves in the $-1$st diffraction order. 
But since $\beta_*$ (of the limiting diffraction solution) is less
than $\pi/L$, we have $\beta_0 \ne - \beta_{-1}$, thus the incoming and
outgoing plane waves have different incident angles. 


In summary, we have found complex modes in an open lossless periodic
waveguide. These modes are physical solutions that can 
be excited whenever the periodic waveguide has a discontinuity or a 
defect.  The complex modes form bands on which the propagation constant
$\beta$ is a complex-valued function of $k$. The bands may have  a 
continuously varying $\mbox{Re}(\beta)$ or a
fixed $\mbox{Re}(\beta)=\pi/L$. At an end of a band, either the 
complex mode turns to a blazing diffraction solution,  or a pair of
complex modes merge to a regular guide mode, or a BIC, or a degenerate
complex mode. Further studies are needed to develop a systematic
approach for computing the complex modes, and to have a deeper
understanding about the complex modes, including the number of bands
and classification of the endpoints. 

\vspace{0.4cm}
\noindent {\bf Funding.} The Research Grants Council of Hong Kong 
Special Administrative Region, China (Grant No. CityU 11305518).



\begin{thebibliography}{99}

\bibitem{snyder} A. W. Snyder and J. Love, {\it Optical Waveguide 
    Theory}  (Springer, US, 1983). 

\bibitem{jablo94} T. F. Jablo\'nski, 
``Complex modes in open lossless dielectric waveguides,'' 
\josaa\ {\bf 11}, 1272--1282 (1994). 

\bibitem{xie11} H. Xie, W. Lu,   and  Y. Y. Lu, 
``Complex modes and instability of full-vectorial beam propagation methods,'' 
\ol\   
{\bf 36}, 2474--2476  (2011). 




\bibitem{mrozo97} M. Mrozowski, {\it Guided Electromagnetic Waves: 
    Properties and Analysis} (Research Studies Press Ltd., England, 
  1997). 


\bibitem{shipman03} S. P. Shipman and S. Venakides, 
 ``Resonance and bound states in photonic crystal slabs,'' 
SIAM J. Appl. Math.  {\bf 64}, 322-342 (2003). 

\bibitem{port05} R. Porter and D. Evans, 
 ``Embedded Rayleigh-Bloch 
   surface waves along periodic rectangular arrays,''  
Wave Motion 
  {\bf 43}, 29-50 (2005). 

\bibitem{mari08} D. C. Marinica, A. G. Borisov, and 
  S. V. Shabanov, 
 ``Bound states in the continuum in photonics,'' 
  \prl\ {\bf 100}, 183902 (2008).   

\bibitem{hsu13_2} C. W. Hsu, B. Zhen, J. Lee, S.-L. Chua, 
  S. G. Johnson, J. D. Joannopoulos, and M. Solja\v{c}i\'{c}, 
  ``Observation of trapped light within the radiation continuum,'' 
  Nature {\bf 499}, 188--191 (2013). 

\bibitem{bulg14b} E. N. Bulgakov and A. F. Sadreev, 
 ``Bloch bound states in the radiation continuum in a periodic array 
 of dielectric rods,'' 
   \pra\ {\bf 90}, 053801 (2014). 

\bibitem{hu15} Z. Hu and Y. Y. Lu, 
 ``Standing waves on two-dimensional 
   periodic dielectric waveguides,'' 
 Journal of Optics {\bf 17}, 
  065601 (2015).  

\bibitem{yuan17} L. Yuan and Y. Y. Lu, 
 ``Propagating Bloch modes above the lightline on a periodic array of 
 cylinders,''  
 J. Phys. B: Atomic, Mol. and Opt. Phys. {\bf 50}, 05LT01 (2017). 


\bibitem{popov01} E. Popov, B. Bozhkov, and M. Nevie\'ere, 
``Almost perfect blazing by photonic crystal rod gratings,'' 
  Applied Optics, {\bf 40}, 2417-2422 (2001). 


\bibitem{fan02} S. Fan and J. D. Joannopoulos, ``Analysis of guided 
  resonances in photonic crystal slabs,'' 
  Phys. Rev. B 
  {\bf 65}, 235112 (2002). 

\bibitem{amgad19} A. Abdrabou and Y. Y. Lu, 
``Indirect link between resonant and guided modes on uniform and 
periodic slabs,'' 
\pra\ {\bf 99}, 063818 (2019). 

\bibitem{gras19} A. Gras, W. Yan, and P. Lalanne, 
``Quasinormal-mode analysis of grating spectra at fixed incidence 
angles,'' 
\ol\
{\bf 44}, 3494-3497 (2019). 

\bibitem{heiss12} W. D. Heiss,  
``The physics of exceptional points,'' 
J. Phys. A: Math. Theor.   {\bf 45}, 444016 (2012). 

\bibitem{zhen15} B. Zhen, C. W. Hsu, Y. Igarashi, L. Lu, I. Kaminer, 
  A. Pick, S.-L. Chua, J. D. Joannopoulos, and M. Soljacic, 
  ``Spawning rings of exceptional points out of Dirac cones,'' 
  Nature (London) {\bf 525}, 354-358 (2015). 

\bibitem{kam17} P. M. Kami\'{n}ski, A. Taghizadeh, O. Breinbjerg, 
  J. Mork, and S. Arslanagi\'{c}, 
  ``Control of exceptional points in photonic crystal slabs,'' 
  Opt. Lett.  {\bf 42}, 2866-2869 (2017). 

\bibitem{amgad18} A. Abdrabou and Y. Y. Lu, 
``Exceptional points of resonant states on a periodic slab,'' 
Phys. Rev. A 
{\bf 97}, 063822 (2018). 

\bibitem{amgad19josab} A. Abdrabou and Y. Y. Lu, 
``Exceptional points for resonant states on parallel circular 
dielectric cylinders,'' 
J. Opt. Soc. Am. B 
{\bf 36}, 1659-1667 (2019). 

\end{thebibliography}
\end{document}